\documentstyle[aps,prb,psfig]{revtex}
\begin{document}

\newcommand {\PRB}       {{\em Phys.\ Rev.\ B}} 
\newcommand {\Vy}        {V$_{2-y}$O$_3$}
\newcommand {\V}         {V$_2$O$_3$}
\newcommand {\be}        {\begin{equation}}
\newcommand {\ee}        {\end{equation}}
\newcommand {\ea}        {{\em et al.}}

\twocolumn[\hsize\textwidth\columnwidth\hsize
\csname @twocolumnfalse\endcsname

\title{Helical spin-density wave in V$_{2-y}$O$_3$}
\author{T. Wolenski, M. Grodzicki, J. Appel\\
Universit\"at Hamburg, 
I. Institut f\"ur Theoretische Physik\\ 
Jungiusstra{\ss}e 9, D-20355 Hamburg.}
\date{Submitted: November 6, 1997}
\maketitle

\begin{abstract}
Recent neutron scattering and nuclear magnetic resonance experiments
have revealed that the low temperature phase of doped \Vy\ is
an itinerant antiferromagnet with a helical spin structure.
We use a band structure calculation as the point of departure to
show that these experiments are in agreement with mean field
results for an Overhauser spin-density wave state. The influences
of a finite life-time and of dilute magnetic impurities are discussed.
\end{abstract}

\vspace{1cm}
]

\narrowtext

\section{Introduction}
\label{sec:intro}

Stoichiometric \V\ is generally considered a classic Mott-Hubbard system
displaying a first order transition at $T_N$=155~K from a paramagnetic
metal (PM) to a local-moment antiferromagnetic insulator (AFI).  
It is common wisdom
that this transition is driven by the strong Coulomb interactions between
conduction band electrons which opens up the Mott-Hubbard gap.  The
antiferromagnetism, which lifts the degeneracy in the insulating phase,
is therefore a consequence of the interaction-driven opening of the 
gap between the upper and lower Hubbard band.

Experimentally the situation appears to be quite different in
doped \Vy, where the low temperature phase is an
antiferromagnetic metal (AFM) with an incommensurate Overhauser
spin-density wave (SDW), similar to metallic chromium.
Neutron scattering experiments \cite{Bao} show that the
SDW has a nesting vector ${\bf Q}=1.70 \, c^*$ 
($c^*$ is a reciprocal lattice vector
in the hexagonal unit cell of \V), and that the transition appears at a 
$T_N$ of about 10~K; both $T_N$ and ${\bf Q}$ are almost independent
of doping in the regime probed. The ordered moment is only
$\mu=0.14 \, \mu_B$ as compared to the $1.2 \, \mu_B$ in the AFI.

Moreover, the magnetic excitations are found\cite{Bao2} to have a bandwidth
of about 20 $k_B T_N$. In a local moment picture one would 
expect a magnetic bandwidth of order $J \sim k_B T_N$. The large bandwidth
can be understood in the framework of SDW theory, where, however,
excitations are found on electronic energy scales.
An interesting observation is that even close to the metal-insulator
transition the magnetic excitations correspond to the SDW spin structure
rather than the magnetic order in the antiferromagnetic Mott insulator
phase. This suggests\cite{WernerPC} that the local vanadium moment 
formation in the AFI phase is accompanied by a change in the 
electronic configuration.

\begin{figure}[tb]
\psfig{file=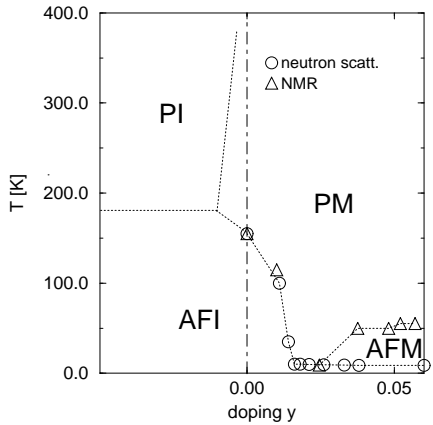,width=9cm}
\vspace{1cm}
\caption{Phase diagram of \Vy\ as derived from neutron scattering,
susceptibility, and NMR measurements. PM: paramagnetic metal, AFI:
antiferromagnetic insulator, AM: antiferromagnetic metal, PI:
paramagnetic insulator.} 
\label{fig:phasediag}
\end{figure}

NMR experiments by Langenbuch and Pieper \cite{Pieper} have 
confirmed the neutron scattering 
results for the ordered moment of
about $0.14 \, \mu_B$ and the orientation of the SDW along
the $c$-direction. The magnitude of the local magnetic field is
identical on all V-sites, compatible with a helical rather than linear 
polarization. At lower doping the temperature dependence of the 
spin-echo intensity also shows $T_N \simeq$10~K.
However, at doping $y \ge 0.038$ the transition
observed with NMR shows $T_N \simeq$ 50~K, with no
special feature in the V relaxation rate around 10~K. 
The phase diagram combining these experimental results is shown in 
Fig.~\ref{fig:phasediag}.
An additional important contribution of NMR is the study of the minority 
V sites neighboring the vanadium vacancies in $c$-direction. 
These sites are approximately in a V$^{4+}$ (S=1) state. 
Furthermore, the 10~K transition could be intrinsic to the \V\ system,
while the magnetic impurities increase the transition temperature
above a certain impurity concentration.
We will discuss this scenario in Sec.~\ref{sec:discussion}.
The homogeneous susceptibility at higher doping \cite{Pieper} 
is very sensitive to
whether the measurement is made in zero field or in the presence
of the earth magnetic field.
In the former case, for $y$=0.052 a transition is seen at 56~K, while
there is no feature at 10~K. In the latter case, there is only some
hysteresis left at higher temperatures, while there is a pronounced peak
at 10~K.

The order of the phase transition is an issue not completely settled 
by these experiments. While the neutron scattering data 
are consistent with a BCS-like second order transition at 
10~K modified by fluctuations,
the NMR data show, at least for the 50~K transition, 
a large discontinuity in the local magnetic field at $T_N$.

There have been attempts to qualitatively study the itinerant AFM
phase using the LISA (local impurity self-consistent
approximation) method starting from the limit of strong
correlations\cite{LISA}, where the key factor for the closing of
the Hubbard gap is the magnetic frustration caused by the defect sites 
and the V$^{4+}$ moments.
However, as pointed out above, the experiments
indicate a conventional Overhauser SDW resulting from 
Coulomb interactions at nested Fermi surface sheets. 
In this paper we show that it is possible to explain
the experiments with a molecular-field Hartree-Fock SDW
calculation starting from accurate band structure results and
taking into account realistic electron-electron scattering
and multiple bands at the Fermi level.

In the following section we discuss the band structure calculation,
which is the point of departure for the Hartree-Fock SDW calculation
with life-time effects discussed in Sec.~\ref{sec:SDW}. The results
will be discussed in Sec.~\ref{sec:results}.
In Sec.~\ref{sec:discussion} we compare our results with the NMR and
neutron scattering experiments and discuss the importance of
magnetic impurities.

\section{Band structure}
The first band structure calculations for \V\ date back to the early
seventies.\cite{band_old}
However, at that time only a tight-binding calculation
with a rather small set of orbitals
was feasible. Recently, Mattheiss \cite{Mattheiss}
performed a state-of-the-art calculation using the scalar-relativistic
linear augmented-plane-wave (LAPW) method. We make use of his results
in the form of a Slater-Koster \cite{SlaterKoster}
calculation using the parameters given by Mattheiss.

In the Slater-Koster method the band structure is constructed
from parameters obtained at special points of high symmetry using LDA-APW 
results. Then, the energy eigenvalues at general ${\bf k}$ are
interpolated using a tight-binding-like formalism, thus preserving
the full symmetry of the lattice. 

To find the energy eigenvalues, in general one has to calculate 
integrals of the form
\be
H_{\mu \mu'}=e^{i {\bf k} ({\bf R}_\mu-{\bf R}_{\mu'})}
\int \phi_\mu^* ({\bf r-R_\mu}) H \phi_{\mu'} ({\bf r-R_{\mu'}}) d{\bf r}
\ee
for a Hamiltonian
\be
H=T+\sum_\kappa V_\kappa ({\bf r-R}_\kappa).
\ee
If $\mu$, $\mu'$ and $\kappa$ are all different, the integrals are
three-center integrals; if $\kappa=\mu$ or $\kappa=\mu'$,
they are two-center integrals.
The three-center integrals tend to be smaller, and the angular
dependence differs only for the $l>0$ contribution, so that it
is a good approximation to only consider two-center integrals
(extended H{\"u}ckel approximation). The wave functions $\phi$ and
the potential $V_\kappa$ are not very well known. 
Therefore, the concept of the Slater-Koster method is to treat the 
two-center integrals as parameters obtained from
LDA calculations at special points, rather than calculate them microscopically.
The remaining non-trivial task is to construct
the full matrix elements between all 44 orbitals considered in the
trigonal unit cell of \V: Five 3$d$-orbitals for each of the four vanadium
atoms and one $2s$- and three $2p$-orbitals for each of 
the six oxygen atoms in the unit cell.

\begin{figure}[tb]
\psfig{file=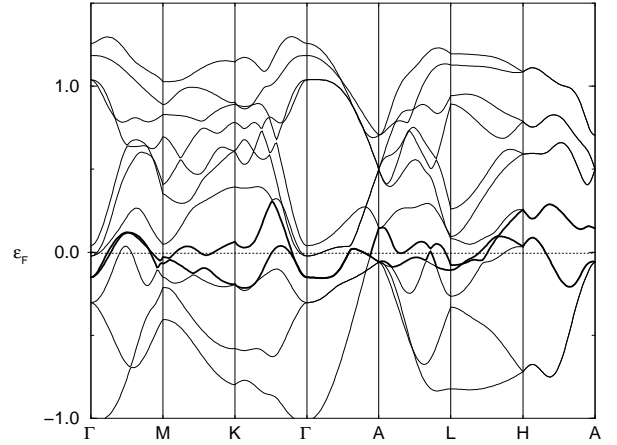,width=9cm}
\caption{Bandstructure in the vicinity of the Fermi energy
($\epsilon=0)$ plotted along lines of high symmetry of the
hexagonal-type Brillouin zone.}
\label{fig:bandstructure}
\end{figure}

\begin{figure}[tb]
\psfig{file=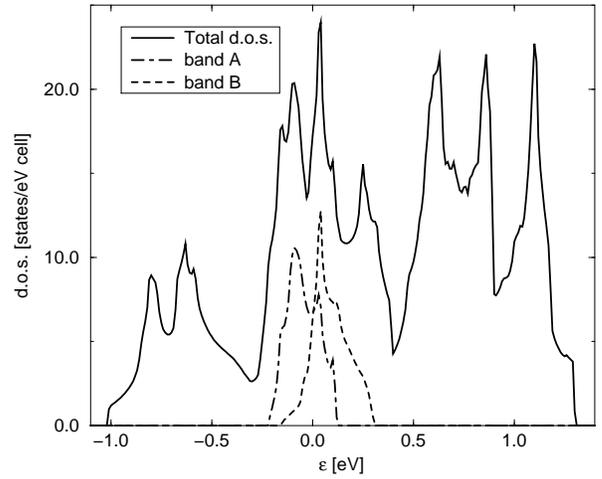,width=9cm}
\caption{Total density of states near the Fermi energy and
partial density of states for the two bands, that contribute most
at the Fermi energy.}
\label{fig:dos}
\end{figure}

In Fig.~\ref{fig:bandstructure} we show the bands in a trigonal crystal
structure plotted along
lines of high symmetry. The symmetry points are labeled using the
hexagonal notation, the Fermi energy is set to be zero. In
Fig.~\ref{fig:dos} 
we plot the density of states in the vicinity
of the Fermi energy, stemming predominantly from the
V(3d)-bands. We also plot the density of states from the
two bands (labeled band A and B respectively)
which contribute about 80 percent 
of the total density of states at
the Fermi energy. Note that the densities of states of band A and band B
at the Fermi energy are almost equal.

\begin{figure}[tb]
\psfig{file=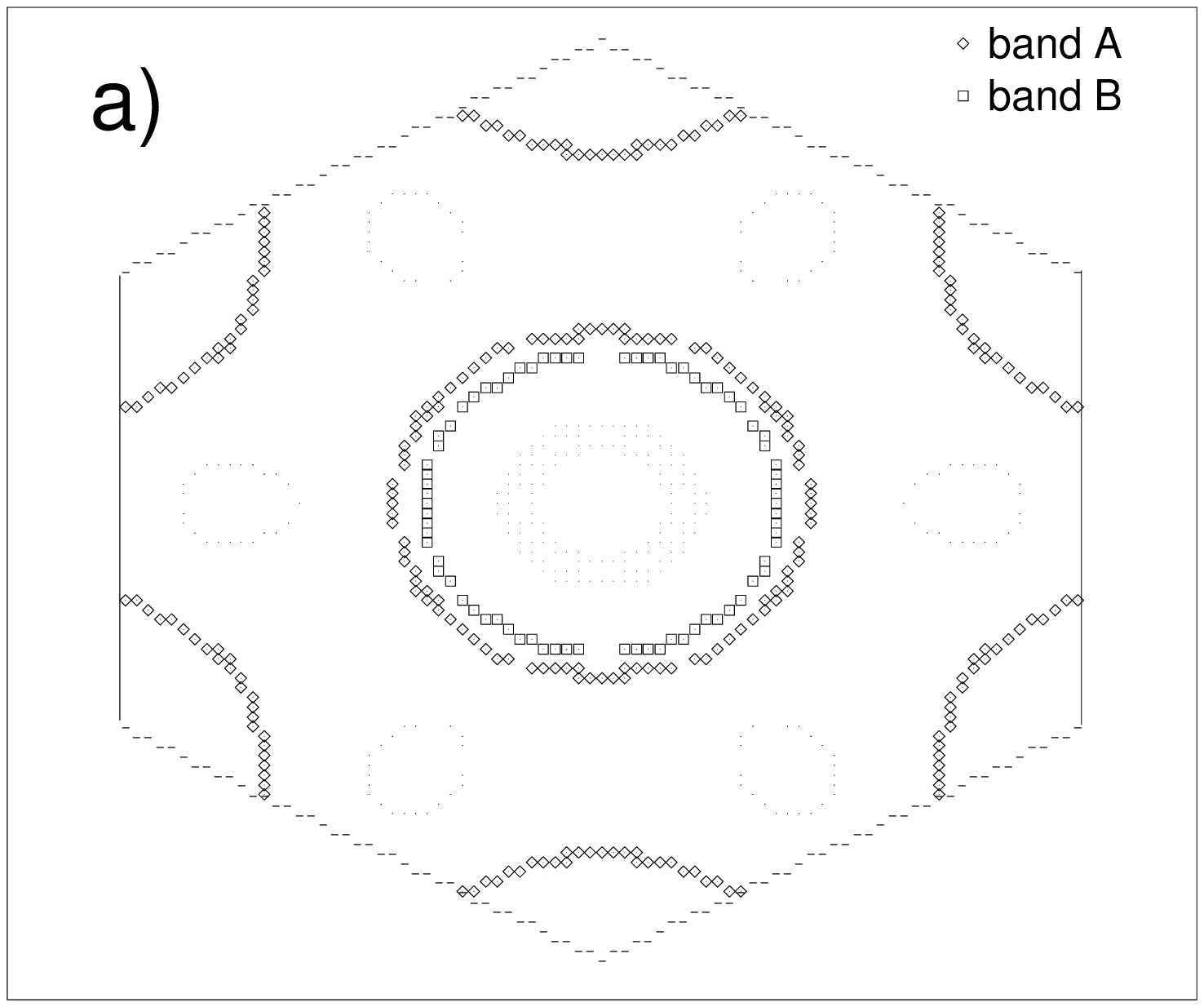,width=9cm}
\vspace{-2cm}
\psfig{file=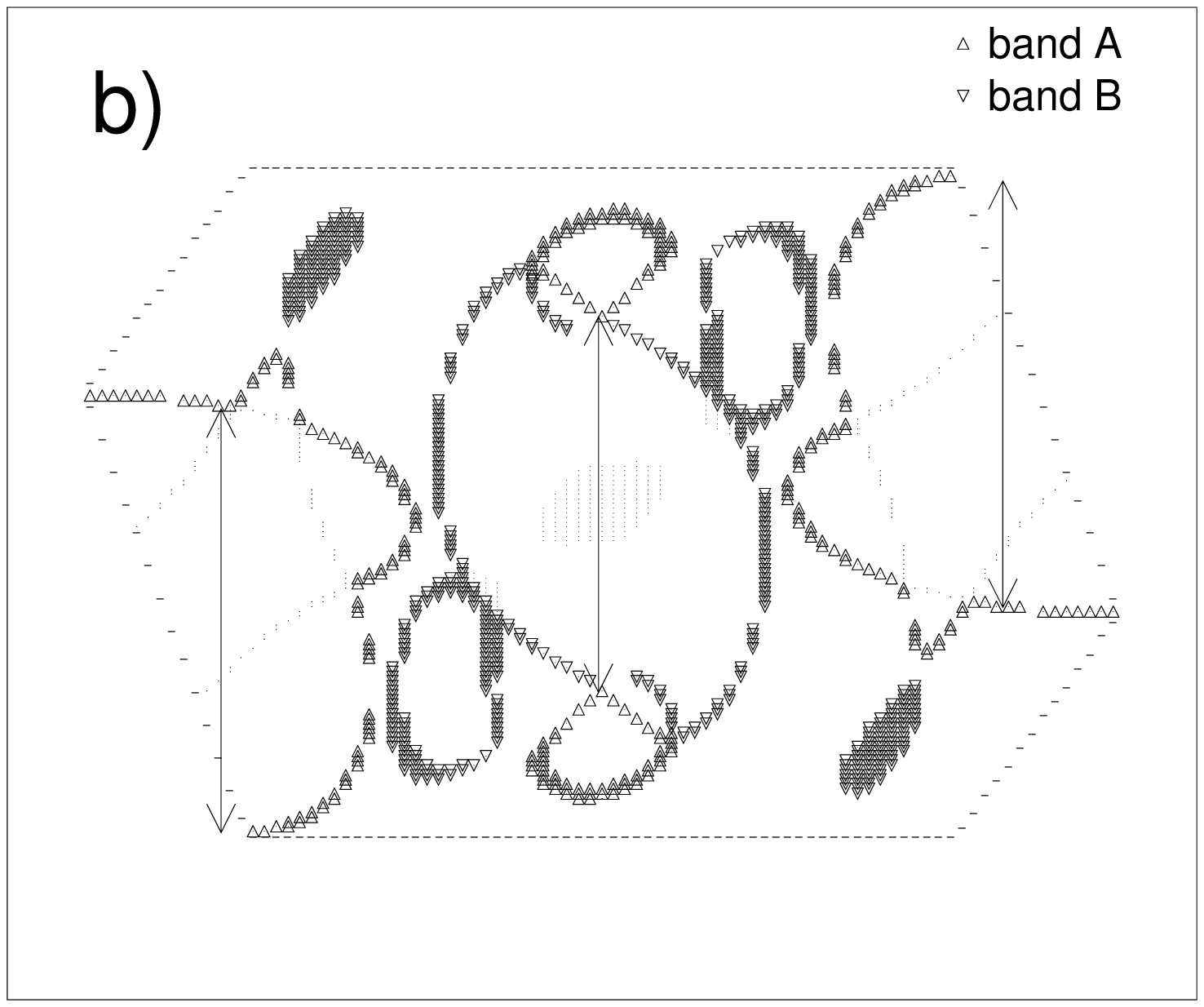,width=9cm}
\caption{Fermi surface in the Brillouin zone of \V. 
a) Plot in the $k_x$-$k_y$-plane ($k_z$=0). 
b) $k_x$-$k_z$-plane ($k_y$=0). The arrows indicate regions of parallel
Fermi surface sheets of band A connected by the numerically determined
nesting vector {\bf Q}.}
\label{fig:fermi}
\end{figure}

The Fermi surface is rather complicated, and therefore we restrict ourselfs
here to showing two cuts through
the Brillouin zone: Fig.~\ref{fig:fermi}~a) 
is in the $a$-$b$ ($k_z$=0) plane  
and reflects nicely the trigonal symmetry of the lattice: 
a three-fold rotation
axis in $c$-direction combined with inversion symmetry around the origin.
This symmetry leads to magnetic frustration in the $a$-$b$ plane,
in agreement with the observed SDW nesting vector in $c$-direction.
In Fig.~\ref{fig:fermi}~b) we show a second cut
in the $a$-$c$ ($k_y$=0) plane, where we indicated the main nesting
feature of the Fermi surface with an arrow. The nesting vector has 
been calculated numerically from the maximum in the susceptibility 
(cf.\ Sec.~\ref{sec:results}), the arrow in this figure only serves for
illustration purposes, in particular since the Fermi surface is
truly three-dimensional and thus a single two-dimensional cut can
be misleading. The nesting part of the Fermi surface we
indicated extends into the $b$-direction, thus giving a sufficient
density of states to produce a spin-density wave as will be shown in the
next section.

Strictly, the band structure calculation is performed for stoichiometric
\V. However, the doping in the samples experimentally probed is 
sufficiently small ($y \le 0.05$) that even at the highest doping levels
the chemical potential is only shifted by a few meV, justifying
a rigid band approximation. 

\section{Helical spin-density wave theory}
\label{sec:SDW}

\subsection{Molecular-field gap equation}

A helical spin-density wave instability, where electrons with 
momentum {\bf k} and holes with momentum {\bf k+Q} and 
opposite spins pair below a critical
temperature and thus produce a gap at parts of the Fermi surface, 
has first been
proposed by Overhauser \cite{Overhauser} for the Hartree-Fock electron
gas. While for the three-dimensional electron gas 
this instability is prevented by screening, certain
Fermi surface topologies with sizeable parallel parts of the Fermi surface
connected by a single vector {\bf Q} (nesting) lead to the appearance of
an SDW phase in actual materials like chromium.\cite{FeddersMartin}

Starting from a Hubbard Hamiltonian
\be
H=\sum_{{\bf k}, \alpha}
\epsilon_{\bf k} c_{{\bf k}, \alpha}^\dagger c_{{\bf k}, \alpha}
+U \sum_{\bf k, k', q} \sum_{\alpha, \alpha'}
c_{{\bf k}, \alpha}^\dagger c_{{\bf k'}, \alpha'}
c_{{\bf k'-q}, \alpha'} c_{{\bf k+q}, \alpha} ,
\ee
the assumption of a finite expectation value for 
$S_{\bf Q}^{\uparrow \downarrow} = \sum_{\bf k} \langle
c_{\bf k+Q \uparrow}^\dagger \sigma^z_{\uparrow \downarrow} 
c_{\bf k \downarrow}
\rangle$
in the Hartree-Fock approximation leads to a molecular-field 
equation for the SDW gap by means of a standard Boguliubov
transformation, rather similar to the BCS gap equation. 
Since the nesting vector {\bf Q} is
determined by the Fermi surface topology, it will in general be
incommensurate with the lattice. The gap equation has been generalized to
the incommensurate situation by Young,\cite{Young} in our notation
it reads
\be
1=\frac{U}{N} \sum_{\bf k} 
\frac{f(E_{\bf k}^+)-f(E_{\bf k}^-)}
{\sqrt{(\epsilon_{\bf k}-\epsilon_{\bf k+Q})^2+4 \Delta^2}}.
\ee
The quasi-particle energies are given by
\be
E^\pm_{\bf k} \equiv \frac{\epsilon_{\bf k}+\epsilon_{\bf k+Q}}{2}
\pm \frac{1}{2} \sqrt{(\epsilon_{\bf k}-\epsilon_{\bf k+Q})^2+4 \Delta^2},
\ee
and $f(E)=1/(\exp(E/k_B T)+1)$ is the Fermi function.

The assumption of a finite expectation value 
$S_{\bf Q}^{\uparrow \downarrow}$ leads to a spiral polarization of the
SDW,\cite{Chubukov} as we can show from considering
$S_{\bf Q}^+ = S_{\bf Q}^{\uparrow \downarrow}\equiv\overline{S}_{\bf Q}$
and
$S_{-\bf Q}^- = S_{-\bf Q}^{\downarrow \uparrow}=\overline{S}_{\bf Q}^*$.
As usual, they are related to $S^x$ and $S^y$ by $S^x=(S^++S^-)/2$ and
$S^y=(S^+ - S^-)/2$. The Fourier transforms then are
\begin{eqnarray}
S_{\bf Q}^x ({\bf R}) & = & 
        (S_{\bf Q}^+({\bf R}) + S_{-\bf Q}^-({\bf R}))/2 \nonumber \\
& = & \overline{S}_{\bf Q} (\exp(i {\bf Q \cdot R}) +
\exp(-i {\bf Q \cdot R}))/2 \nonumber \\
& = & \overline{S}_{\bf Q} \cos {\bf Q \cdot R} \\
S_{\bf Q}^y ({\bf R}) & = & \overline{S}_{\bf Q} \sin{\bf Q \cdot R},
\end{eqnarray}
which is a spiral polarization. If the nesting instability were 
strong enough to simultaneously support an SDW with opposite
helicity, we would get a linear polarization. In the
present case, however, the subtle modification of the Fermi surface 
through the formation of one spiral SDW is sufficient to prevent this
additional, opposite SDW.

\subsection{Life-time effects due to quasiparticle scattering}

The analysis we presented so far is for a one-band model without
any scattering. As we show in the preceding section, at least two
bands are relevant for the low energy physics of \V. Testing the
nesting properties of these two bands, it shows however, that only
one of them shows significant nesting leading to a possible
SDW instability. Hence, our approach is to assume that the SDW results
from the nesting band A, while band B
enters the physics via scattering processes from the part of the inter-band
Coulomb interaction not already taken into account in the LDA band
structure.
Since the densities of states at the Fermi energy of bands A and B
are comparable, this scattering is expected to significantly shorten
the life-time of band A quasi-particles.

The scattering results from both electron-electron 
interactions, $\tau_{\text{e-e}}^{-1}$, and the 
scattering of electrons from the 
vanadium vacancy sites, $\tau_{\text{imp}}^{-1}$. The sizeable
electron-electron scattering, which has the form\cite{AppelOverh} 
($N_A$ and $N_B$ are
the densities of states for the respective bands)
\be
\tau_{\text{e-e}}^{-1} = c N_A(0) N_B(0) (\pi^2(k_B T)^2+(\hbar \omega)^2),
\ee
is responsible for the $T^2$ term in the electrical 
resistivity,\cite{Carter} while $\tau_{\text{imp}}^{-1}$ 
determines the residual resistivity at $T=0$.
We thus have
\be
\tau^{-1} = c N_A(0) N_B(0) (\pi^2(k_B T)^2+(\hbar \omega)^2) + 
\tau_{\text{imp}}^{-1}.
\ee
The constant $c$ contains the scattering cross section. To estimate
$c$ we use the relation between resistivity and scattering rate,
\be
\rho=\frac{m^*}{n e^2} \, \tau^{-1}.
\ee
The carrier density $n$ can be estimated from the Hall coefficient,
\be
R_H = \frac{1}{ne}.
\ee
Experimentally,\cite{McWhan} $R_H = 3.5 \times 10^{-4}$ cm$^{3}$ $C^{-1}$,
resulting in a carrier density $n=1.8 \times 10^{22} \mbox{cm}^3$.
The coefficient of the $T^2$ term in the resistivity\cite{Carter} is
$6.0 \times 10^{-8} \, \Omega \, \mbox{cm K}^{-2}.$
Taking all these values derived from different experiments at face
value, a realistic effective mass $m^*$ of a few times $m$ 
results in a scattering rate $\hbar \tau^{-1} / k_B \simeq$ 10~K
at $T_N$.
One should consider this an estimate of $\tau^{-1}$, which serves to prove the
consistency of our approach, rather than an
exact value, and hence we will parameterize
the scattering rate by multiplying with a scaling parameter $\alpha$,
which varies between 0.1 and 1, i.e.
\begin{equation}
\hbar \tau = \alpha (k_B T)^2.
\end{equation} 

To take these life-time effects into account we make use of an
approach first applied to the case of BCS superconductors with
strong damping of the quasi-particles by Bardasis and
Schrieffer.\cite{Bardasis}
These authors derive the molecular-field gap equation including the
Nambu-Eliashberg function $Z({\bf q}, \omega)$, which is related
to the scattering rate $\tau^{-1}$ via
\be
Z({\bf k}, \omega)=\mbox{Re} Z({\bf k}, \omega)
     \left( 1+ i \frac{\tau^{-1}}{\omega} \right).
\ee
The real part of $Z({\bf q}, \omega)$ renormalizes the band structure
and is already taken into account in LDA. Thus, we set 
$\mbox{Re} Z({\bf k}, \omega)=1$.
The imaginary part of $Z({\bf q}, \omega)$ is taken into account by
 straightforward generalization of the gap equation of 
Ref.~\onlinecite{Bardasis} to the case of an incommensurate
SDW and to finite temperatures by employing the usual 
summation technique over Matsubara frequencies;
one arrives at
\begin{eqnarray}
\label{eq:gap}
1=\frac{U}{N} \mbox{Re} \sum_{\bf k} 
\frac{f(E_{\bf k}^+)-f(E_{\bf k}^-)}
{Z \sqrt{(\epsilon_{\bf k}-\epsilon_{\bf k+Q})^2+4 \Delta^2}} \times \nonumber \\
\times  \left( \left. 1-E_{\bf k} \frac{d}{d\omega} \frac{1}{Z} 
\right|_{\omega=E_{\bf k}/Z} \right)^{-1}.
\end{eqnarray}

\section{Results}
\label{sec:results}
\begin{figure}[tb]
\psfig{file=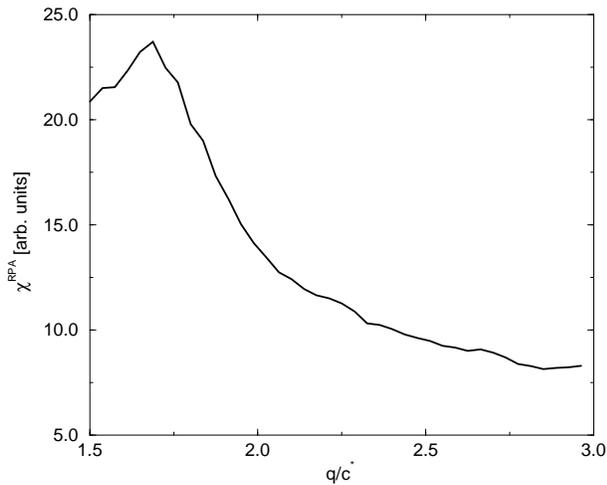,width=9cm}
\caption{The RPA susceptibility for $U=0.4$~eV without any quasi-particle
scattering in the paramagnetic state just above the SDW transition.
Other values of $q/c$ are accessed using the periodicity and the 
inversion symmetry of the lattice.}
\label{fig:chi}
\end{figure}

The necessary first step is to determine the nesting vector ${\bf Q}$.
There are two basically equivalent ways to do this: One can
examine the $\bf q$-dependent susceptibility
\be
\chi^0({\bf q})=\sum_{\bf k} \frac{f(\epsilon_{\bf k+q})-f(\epsilon_{\bf k})}
{\epsilon_{\bf k+q}-\epsilon_{\bf k}}
\ee
and look for a maximum at a position ${\bf q}_{\text{max}}$.
Within RPA, where the full susceptibility
is given by 
$\chi_{\text{RPA}}({\bf q})=\chi^0({\bf q})/(1-U \chi^0({\bf q}))$,
with increasing $U$ the phase transition will first occur for
${\bf Q=q}_{\text{max}}$. Equivalently, one can study the gap equation
Eq.~(\ref{eq:gap}), and calculate the value of ${\bf q}$ that leads to
the minimum value of $U$ necessary to satisfy the gap equation at a given
temperature, i.e. to produce the SDW phase transition. 
Both methods lead to the same
result ${\bf Q}=(1.69 \pm 0.03) \, c^*$ in surprisingly
good agreement with the neutron scattering results.\cite{Bao}
In Fig.~\ref{fig:chi} 
we show the susceptibility plotted for
different values of ${\bf q}$ along the $c$-direction 
in the paramagnetic state close to the SDW transition.

\begin{figure}[tb]
\psfig{file=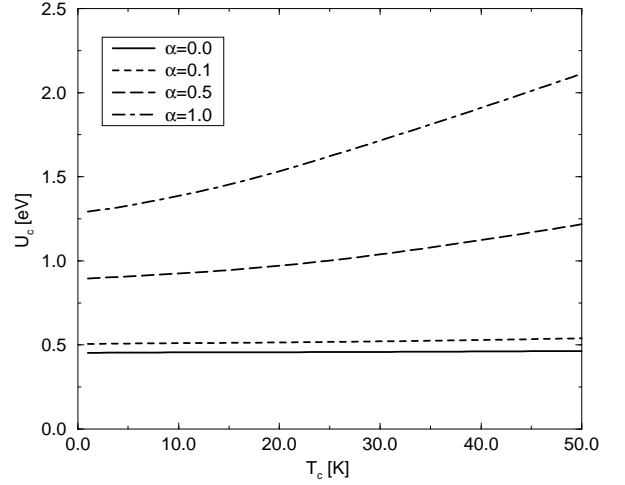,width=9cm}
\caption{Critical interaction necessary to achieve a given $T_N$.
The electron-electron parameter $\alpha$ determining 
the scattering strength is explained in the text.}
\label{fig:Uc_von_Tc}
\end{figure}

Since the effective screened interaction $U$ is not known very well from 
first principles, we use it as a parameter to fix $T_N$. 
From optical conductivity data\cite{Thomas}
$U$ has been estimated to be about 1 eV, a value also used in
the LISA calculations \cite{LISA}.
In Fig.~\ref{fig:Uc_von_Tc}
we plot the interaction $U_c$ necessary to obtain a given
critical temperature for the SDW transition. 

\begin{figure}[tb]
\psfig{file=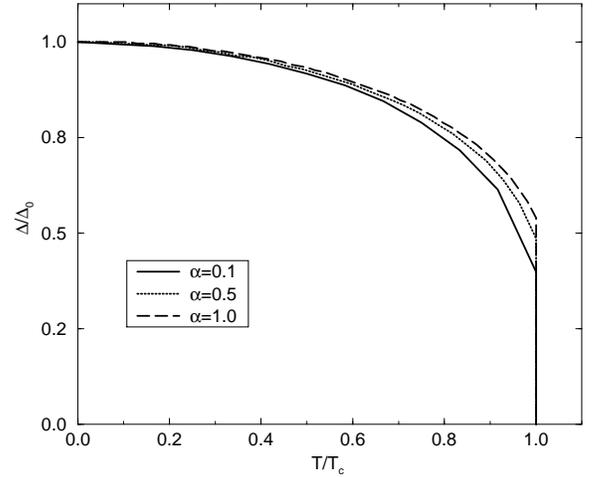,width=9cm}
\caption{Order parameter $\Delta(T)$ plotted against reduced temperature
$T/T_N$ for different strengths of quasi-particle scattering.}
\label{fig:gap}
\end{figure}

The next step is to solve the gap equation for given $U$ and $\tau$
for the SDW gap $\Delta(T)$. In Fig.~\ref{fig:gap} 
we show the results
for a $T_N$ of 10~K and different scattering rates. Clearly, the jump
in the order parameter at $T_N$ increases with increasing scattering rate.
However, even at the lower limit of the reasonable range of scattering
rates this discontinuity is significant, in agreement with 
the findings of Pieper {\em et al.}\cite{Pieper}.

There has been some discussion in the context of chromium 
as to how a mean-field SDW transition can become first order.
One suggestion is the effect of a ``reservoir band'',\cite{Rice}
which is polarized by the formation of the SDW. Clearly, a
strong coupling to the lattice will also enhance the tendency
towards a first order transition. In the present case, however, 
the strong lifetime effects of our two-band model already produce
a first order transition. The life-time effects are stronger for
the 50~K transition than for the 10~K transition, so that it is possible
that the 10~K transition is modified once fluctuations are taken 
into account.  

It is hard to estimate the size of the ordered magnetic moment from
this calculation, since we did not test the full ${\bf q}$-dependence
of the order parameter. However, we calculated that about ten
percent of the Fermi surface make the main contribution to the 
SDW instability, in agreement with an estimate of Bao {\em et al.}
\cite{Bao}.

\section{Discussion and Conclusions}
\label{sec:discussion}

Taking an LDA band structure calculation as point of departure, we
find a SDW transition at 10~K with realistic parameters with
a nesting vector ${\bf q}$=1.69 $c^*$ and a helical spin structure
in agreement with neutron scattering results.\cite{Bao}
We believe that this SDW is the intrinsic low temperature 
phase of doped \Vy.
This leaves the puzzle of the partial disagreement of neutron scattering
and NMR\cite{Pieper} as to what the transition temperature of the
higher doped compounds is. 

We suggest the following scenario for the higher doped compounds. 
The V site neighboring a vanadium vacancy in $c$-direction is in
a V$^{4+}$ state\cite{Pieper} with $\mu=1 \mu_B$. 
These moments are very dilute at the relevant 
concentrations $y \le .05$. Overhauser\cite{Overhauser2} and recently
Werner\cite{Werner} have shown how an incipient SDW combined with 
magnetic impurities can lead to an SDW transition. The SDW is driven
by the maximization of the interaction energy 
$E_{\text{int}}={\bf H}_j \cdot {\bf S}_j$,
where ${\bf H}_j$ is the local field due to the SDW and ${\bf S}_j$ the
impurity spin. If the RKKY-type interactions between the dilute
impurities are strong enough, this mechanism will also lead to the formation
of a domain structure, which could explain the difference between 
neutron scattering and the more local NMR experiments.

An open question is whether the delicate field dependence in the
experiments of Pieper {\em et al.}\cite{Pieper} is related to an
ordering of the V$^{4+}$ moments, preventing the formation of helical
SDW domains. Problematic is that the critical field, which would have to
overcome the RKKY interaction between the 
dilute moments, is only of the order of 1~G. 
It is known that strong non-linearities exist in spin glass 
systems,\cite{Binder}
where comparable fields are sufficient to suppress a sharp cusp in the 
susceptibility. However, this point clearly requires further attention,
to clarify the experimental findings.

Another open question is how the Mott transition at 155~K, which apparently is
accompanied by a change in the electronic configuration---and also in
the lattice symmetry, which becomes monoclinic in the AFI 
insulator---is influenced by taking into account the realistic
multi-band bandstructure. Particularly the intriguing experimental
observation, that the fluctuations on the metallic side of the
metal-insulator type are of the SDW rather than local moment type
with a magnetic structure given by Fermi surface topology, 
show that for the physics of the Mott transition in \V\
both strong correlation and electronic structure effects
are important. This problem is under investigation.

\section*{Acknowledgments}
We like to thank M. W. Pieper for insightful discussions. 
T. W. acknowledges financial support through a fellowship from 
the Universit\"at Hamburg. 


\end{document}